\documentclass{aa}
\usepackage{graphicx}
\usepackage{subfig}

\usepackage{txfonts}

\begin{document}


\title{Distribution of methanol and cyclopropenylidene around starless cores \thanks{Based on observations carried out with the IRAM 30m Telescope. IRAM is supported by INSU/CNRS (France), MPG (Germany), and IGN (Spain)}
}


\author{S. Spezzano\inst{\ref{inst1}} \and P. Caselli\inst{\ref{inst1}} \and J. E. Pineda \inst{\ref{inst1}} \and L. Bizzocchi \inst{\ref{inst1}} \and D. Prudenzano \inst{\ref{inst1}} \and Z. Nagy\inst{\ref{inst1}}$^,$\inst{\ref{inst2}}}

\institute{Max Planck Institute for Extraterrestrial Physics, Giessenbachstrasse 1, 85748 Garching, Germany \label{inst1}
\and
Konkoly Observatory, Research Centre for Astronomy and Earth Sciences, H-1121 Budapest, Konkoly Thege\'ut 15--17, Hungary \label{inst2}
}
 
\abstract {The spatial distribution of molecules around starless cores is a powerful tool for studying the physics and chemistry governing the earliest stages of star formation.}{Our aim is  to study the chemical differentiation in starless cores to determine the influence of large-scale effects on the spatial distribution of molecules within the cores. Furthermore, we want to put observational constraints on the mechanisms responsible in starless cores for the desorption of methanol from the surface of dust grains where it is efficiently produced.}{We mapped methanol, CH$_3$OH, and cyclopropenylidene, $c$-C$_3$H$_2$, with the IRAM 30m telescope in the 3 mm band towards six starless cores embedded in different environments, and in different evolutionary stages. Furthermore, we  searched for correlations among physical properties of the cores and the methanol distribution. }{From our maps we can infer that the chemical segregation between CH$_3$OH and $c$-C$_3$H$_2$ is driven by uneven illumination from the interstellar radiation field (ISRF). The side of the core that is more illuminated has more C atoms in the gas-phase and the formation of carbon-chain molecules like $c$-C$_3$H$_2$ is enhanced. Instead, on the side that is less exposed to the ISRF the C atoms are mostly locked in carbon monoxide, CO, the precursor of methanol.}
{We conclude that large-scale effects have a direct impact on the chemical segregation that we can observe at core scale. However, the non-thermal mechanisms responsible for the desorption of methanol in starless cores do not show any dependency on the H$_2$ column density at the methanol peak. }
\keywords{ISM: clouds - ISM: molecules - radio lines: ISM
               }
\maketitle

\section{Introduction}
Stars form in dense cloud cores that can be studied through the emission of cold dust and molecules. Molecules, in particular, carry very precious information, and studying the chemical structure of dense cores is important to reveal the process of star formation and the chemical evolution from clouds to stars and planets.
Starless cores are the earliest stages in the formation of low-mass stars. The more dynamically evolved cores among starless cores are the pre-stellar cores, gravitationally unstable cores on the verge of collapse to form a protostar. 
The key evolutionary factors that define a pre-stellar core are extensively described in \cite{Crapsi05}, among which are the high abundance of N$_2$H$^+$ and N$_2$D$^+$, high deuterium fractionation, high CO freeze-out towards the centre of the core, and central density higher than 5$\times$10$^5$ cm$^{-3}$ (see also \citealt{Keto08}).
Small-scale effects influence the chemical differentiation in starless cores. The enhanced deuteration observed in the inner regions of starless cores, for example, is caused by the local physical properties of the centre of starless cores: low temperatures (5-10 K) and high densities (10$^5$-10$^6$ cm$^{-3}$) \citep{Caselli03}. 
\newline The chemical segregation of methanol, CH$_3$OH, and cyclopropenylidene, $c$-C$_3$H$_2$, towards the pre-stellar core L1544 is discussed in \cite{Spezzano16}. The different spatial distribution of  $c$-C$_3$H$_2$ and CH$_3$OH can be related to the uneven illumination around the core, a large-scale effect.  While $c$-C$_3$H$_2$ traces the southern part of the core, where a sharp drop in the H$_2$ column density is present, CH$_3$OH traces the northern part of the core, characterised by a shallower tail of H$_2$ column density. This differentiation can be related to the fraction of carbon locked in carbon monoxide, CO. In the south of L1544, photochemistry maintains more C atoms in the gas phase, and consequently increases the production of carbon-chain molecules. In the north of L1544, C is mainly locked in CO, the precursor for the formation of methanol on dust grain. The emission maps of 22 different molecules were observed towards the same core, and it was shown that all oxygen-bearing molecules have their emission peak towards the  north of the core, and show a similar distribution to methanol \citep{Spezzano17}. In order to test the universality of the above-mentioned results on L1544, we decided to study the chemical differentiation in six starless cores in different evolutionary stages, and embedded in different environments. 

Furthermore, studying the distribution of methanol around starless cores is essential to understand the physical and chemical processes, still poorly understood, that are responsible for its desorption from the icy mantles of dust grains. Methanol is the simplest complex organic molecule (COM, defined as carbon-containing molecules with more than five atoms, \citealt{herbst09})observed in the ISM, and the one whose formation reactions are better studied and constrained. Gas-phase reactions fail to reproduce the high abundances of methanol observed towards hot cores and corinos \citep{Garrod&Herbst06,Geppert06}. On the other hand, laboratory experiments show that methanol can be efficiently formed through the hydrogenation of carbon monoxide frozen on the icy mantle of dust grains \citep{Watanabe02}. Hot core and hot corinos are high temperature regions (T$\geq$100 K) surrounding high- and low-mass protostars, respectively. The observation of large fractional abundances of methanol relative to molecular hydrogen towards hot cores and hot corinos can be explained with methanol evaporating from the grains when the medium is warmed up by the protostar. Explaining the large fraction of methanol observed around starless cores is more problematic because the mechanism of thermal desorption from dust grains must be ruled out due to the absence of a nearby heating source. Recent laboratory works show that photodesorption of methanol from dust grains cannot be the mechanism responsible for the gas-phase methanol observed around starless cores \citep{Bertin16,Cruz16} as the methanol will mostly desorb as fragments and not as a whole. Alternative routes for an efficient desorption of methanol from the icy mantles at low temperatures are the reactive or chemical desorption \citep{Vasyunin17}, low-velocity shocks caused by accretion of material onto the core \citep{Punanova18}, and sputtering due to cosmic rays \citep{Dartois19}. Because methanol is the starting point of chemical complexity in the ISM, understanding the processes responsible for its release in the gas phase is of pivotal importance in order to constrain chemical models and reproduce the chemical complexity that we observe in space.

This paper is structured as follows: the source sample and the observations are described in Sect. 2, the chemical segregation observed between $c$-C$_3$H$_2$ and CH$_3$OH is described in Sect. 3, the abundance profiles of methanol and cyclopropenylidene are presented in Sect.4, a discussion about methanol in starless cores is presented in Sect. 5. In Sect. 6 we summarise our results and present our  conclusions.

\section{Observations}\label{Observations}
\subsection{Source sample}
Our source sample, listed in Table~\ref{table:sources},   covers a wide range of evolutionary stages and environmental conditions.
Among the six cores mapped, four are considered   pre-stellar following the definition given in \cite{Crapsi05}, namely L429, L694-2, OphD, and HMM-1. 
B68 is an isolated starless core, called a Bok globule, located in the south with respect to the Ophiucus molecular cloud. Depletion of N$_2$H$^+$ and C$^{18}$O has been observed towards this source \citep{Bergin02}. B68 was included in our study because it is relatively isolated, and hence supposed to be exposed to uniform external illumination. Therefore, we expect the segregation among $c$-C$_3$H$_2$ and CH$_3$OH to be minimised.
L429, in the Aquila Rift molecular cloud, is the core with the highest H$_2$ column density in our sample and is characterised by a very high deuterium fraction in N$_2$H$^+$ \citep{Crapsi05}. 
L694-2 is a pre-stellar core with structural and chemical properties similar to L1544 \citep{Lee04,Crapsi05}.
L1521E is a starless core in Taurus, the same molecular cloud where L1544 is located, but it is not as evolved  \citep{Tafalla04, nagy19}.
OphD and HMM-1 are two pre-stellar cores in Ophiucus. 
B68, L694-2, and L1521E are embedded in a homogeneous interstellar radiation field (ISRF), and hence the illumination on the core only depends on the density structure around it. B68 is a particular case because it is an isolated core and as a consequence an even illumination can be assumed.
L429, OphD, and HMM-1 are instead embedded in active environments \citep{Redaelli18,deGeus92,Preibisch08}, and   the presence of  nearby stars, as well as the density structure around the core, has an influence on the illumination on it. Our aim is to check whether large-scale effects, like  illumination, can indeed be responsible for the chemical differentiation that we observe at core scales.

\subsection{IRAM 30m}
The maps presented in this paper, shown in Figures~\ref{fig:maps} and ~\ref{fig:appendix}, were observed using the Eight MIxer Receiver (EMIR) E090 instrument of the IRAM 30m telescope (Pico Veleta) in on-the-fly mode with position switching. The off positions are 3$\arcmin$  from the dust peak. The molecular transitions, observed in a single set-up, are reported in Table~\ref{table:parameters}. The Fourier transform spectrometer (FTS) was used as the backend with a spectral channel resolution of 50 kHz, about 0.15 km s$^{-1}$ in the 3 mm frequency range. The observations were carried out in June and August 2016. The antenna moved along an orthogonal pattern of linear paths separated by 8$\arcsec$ intervals, corresponding to roughly one-third of the beam full width at half maximum (FWHM). The mapping was carried out in good weather conditions ($\tau$ $\sim$ 0.03) and a typical system temperature of T$_{sys}$ $\sim$ 90-100 K.  A single set-up was used, and the typical noise level per spectral channel resolution is between 15 and 30 mK km s$^{-1}$. The data processing was done using the GILDAS software \citep{Pety05}. A typical second-order polynomial baseline was used for all the sources. All the emission maps presented in this paper are gridded to a pixel size of 4$\arcsec$ with the CLASS software in the GILDAS package, which corresponds to 1/5-1/7 of the actual beam size, depending on the frequency. The velocity ranges used to calculate the integrated intensities are reported in Table~\ref{table:sources}.

\begin{table*}
\centering
\caption{Sources }
\label{table:sources}
\scalebox{0.8}{
\begin{tabular}{ccccccccc}
\hline\hline
&Right Ascension & Declination &V$_{LSR}$ &Distance\tablefootmark{a}& H$_2$ Column Density\tablefootmark{b}  &  Velocity range\tablefootmark{c}&Map size 30m&\\
&(J2000)&(J2000)&(km s$^{-1}$)&(pc)& (10$^{22}$ cm$^{-2}$) &(km s$^{-1}$)& ($\arcsec\times\arcsec$) \\
\hline \hline
L1521E (s)\tablefootmark{d}&04:29:15.7&+26:14:05.0&6.7&140& 2.5 &6.2-7.3  &70$\times$70\\
HMM-1 (p)&16:27:58.3&$-$24:33:42.0&4.3&140& 6 & 3.7-4.6 &75$\times$85\\
OphD (p)&16:28:30.4&$-$24:18:29&3.5&140& 4 &3.2-4.0&85$\times$85 \\
B68 (s)&17:22:38.9 & $-$23:49:46&3.4&150& 1.4& 2.9-3.7  &80$\times$80 \\
L429 (p)&18:17:6.40&$-$08:14:00&6.7&436& 10 & 6.2-7.4 &75$\times$85\\
L694-2 (p)& 19:41:04.5&+10:57:02&9.6&230& 8 &9.2-10 &70$\times$70\\
\hline
\end{tabular}}
\tablefoot{
\tablefoottext{a}{B68: \cite{Alves2007}, L429, OphD and HMM-1: \cite{Ortiz2018}, L694-2: \cite{Kawamura2001}, L1521E: \cite{Galli2018}}
\tablefoottext{b}{Value computed from {\em Herschel}/SPIRE observations towards the dust peak.}
\tablefoottext{c}{Velocity ranges where the integrated emission has been computed.}
\tablefoottext{d}{starless (s) or pre-stellar (p), following the definition given in \cite{Crapsi05}}}
\end{table*}

\subsection{{\em Herschel}/SPIRE}
The large-scale maps presented in this paper were dowloaded from the ESA {\em Herschel} Science Archive\footnote{http://archives.esac.esa.int/hsa/whsa/}. We computed the total column density of H$_2$, N(H$_2$), and the dust temperature, T$_{dust}$, using the three Spectral and Photometric Imaging Receiver (SPIRE) bands at 250 $\mu$m, 350 $\mu$m, and 500 $\mu$m, for which the pipeline reduction includes zero-level corrections based on comparison with the Planck satellite data \citep{Griffin10}. The maps of  N(H$_2$) and T$_{dust}$ are shown in Figures~\ref{fig:H2maps} and \ref{fig:Tdust}. A modified blackbody function with the dust emissivity spectral index $\beta$ = 2.0 was fitted to each pixel, after smoothing the 250 $\mu$m and 350 $\mu$m maps to the resolution of the 500 $\mu$m image ($\sim$40$\arcsec$) and resampling all images to the same grid. For the dust emissivity coefficient we adopted the value from \cite{Hildebrand83}, $\kappa$$_{250}$ $\mu$m = 0.1 cm$^ 2$ g$^{-1}$, and used the power law approximation\\
\begin{eqnarray*}
\kappa_{\nu} &=& \kappa_{\nu_0}\left(\frac{\nu}{\nu_0}\right)^{\beta},
\end{eqnarray*}
\noindent where $\kappa_{\nu_0}$ is the opacity at a reference frequency 250 $\mu$m. A dust-to-gas mass ratio of 0.01 is assumed.
The N(H$_2$) peak value for each source is listed in Table~\ref{table:sources}. All {\em Herschel} maps presented in this paper have a size of 9$\times$9 arcmin$^2$. The T$_{dust}$ maps are in the Appendix.

\section{Results}\label{results1}
We present here the maps of methanol ($J_{K_a,K_c}$ = 2$_{1,2}$-1$_{1,1}$, $E_2$) and cyclopropenylidene ($J_{K_a,K_c}$ = $2_{0,2} - 1_{0,1}$) towards two starless and four pre-stellar cores with the IRAM 30m telescope (Figure~\ref{fig:maps}). A segregation between the two molecules is present towards all cores in our sample except for B68. Given the broadband capabilities of the receivers at the IRAM 30m telescope, we were able to map several molecules simultaneously. In \cite{Spezzano17} it was shown that towards the prototypical pre-stellar core L1544 all carbon-chain molecules trace the same part of the core (the $c$-C$_3$H$_2$ peak), while oxygen-bearing molecules peak towards a different part of the pre-stellar core (the methanol peak). Sulphur does not seem to play a key role in the spatial distribution, with OCS and SO peaking towards the methanol peak and CCS and H$_2$CS peaking at the $c$-C$_3$H$_2$ peak. Based on previous studies of L1544 \citep{Spezzano16, Spezzano17}, we   decided to use, when available, the maps of CCS and C$_4$H to check if the emission structure of cyclopropenylidene, of which we have mapped only one transition, is indeed representative of the distribution of carbon chain molecules. 
For the methanol line, we use instead the $2_{0,2} - 1_{0,1}$ (E$_1$) transition, when available, because it is slightly higher in energy than the 2$_{1,2}$-1$_{1,1}$ ($E_2$) transition, and hence less optically thick. These additional maps are reported in Appendix A, Figure~\ref{fig:appendix}.
Given the frequency resolution of our dataset, 50 kHz, we cannot exclude the effect of self-absorption on the maps, nor  did we observe enough lines of methanol and cyclopropenylidene to be able to precisely derive the T$_{ex}$, and hence the optical depth of the mapped transitions. Assuming an excitation temperature of 5 K, we  calculated the optical depth of the mapped transitions towards the dust peak in each source (see Table~\ref{table:tau}).
The line opacity $\tau$ is defined as

\begin{equation}
\tau = \ln\left(\frac{J(T_{ex}) -  J(T_{bg})}{J(T_{ex}) - J(T_{bg}) - T_{mb}}\right), \label{eq:1}\\
\end{equation}

\noindent where $J(T) = {\frac{h\nu}{k}}(e^{\frac{h\nu}{kT}}-1)^{-1}$ is the equivalent Rayleigh-Jeans temperature in Kelvin, k is the Boltzmann constant, $\nu$
is the frequency of the line, $h$ is the Planck constant, and $T_{ex}$, $T_{bg}$, $T_{mb}$ are the excitation, the
background (2.7 K), and the main beam temperatures respectively, in K. 
We  assume an excitation temperature of 5 K for all levels \citep{caselli02}. This temperature assumption gives a $\tau$ of 0.87 for the lines of $c$-C$_3$H$_2$ towards L1521E, and of methanol towards HMM-1 and L429, showing moderately high optical depth.
 The properties of all observed transitions are given in Table~\ref{table:parameters}. The critical densities of the two methanol transitions are calculated based on the Einstein coefficients and collisional rates in the LAMDA catalogue \citep{LAMDA}; instead,  for the cyclopropenylidene transitions  we   use the collisional rates reported in \cite{Kalifa2019}. The critical densities are not reported for C$_4$H and CCS because the collisional coefficients are not present in the LAMDA catalogue. Furthermore, in order to check whether the illumination around the core has an influence on the distribution that we observe, we   plot the contours of both molecules on the H$_2$ column density and T$_{dust}$ maps that we computed from the {\em Herschel}/SPIRE observations (Figure~\ref{fig:H2maps}). To facilitate the comparison among the different cores, the contours from the SPIRE maps are reported in visual extinction, calculated as A$_V$ [mag] = N(H$_2$)[cm$^{-2}$]/9.4$\times$10$^{20}$ [cm$^{-2}$ mag$^{-1}$] \citep{Bohlin78}. The spectra of methanol and cyclopropenylidene extracted at the dust peak of each of the sources in our sample are shown in Figure~\ref{fig:spectra}. All lines are fitted with a Gaussian profile, and the corresponding parameters are given in Table~\ref{table:lines}. Given the low spectral resolution of the observations ($\sim$0.15 km s$^{-1}$), we cannot infer detailed information about the kinematics of the sources in our sample. Nevertheless, we can confirm the same results reported in \cite{Spezzano16} for methanol and cyclopropenylidene in L1544: despite the different spatial distribution, both molecules trace   similar kinematic patterns in all sources in our sample, and have similar line widths.

\begin{table}
\centering
\caption{Optical depth of the mapped lines at 5 K}
\label{table:tau}
\scalebox{1}{
\begin{tabular}{c|cc}
\hline\hline
&$c$-C$_3$H$_2$ &CH$_3$OH  \\
&($2_{0,2} - 1_{0,1}$)&(2$_{1,2}$-1$_{1,1}$, $E_2$)\\
\hline
L1521E&0.87(5)&0.62(4)\\
HMM-1&0.34(3)&0.87(8)\\
OphD&0.62(4)&0.38(3)\\
B68&0.7(1)&0.26(4)\\
L429&0.57(4)&0.87(6)\\
L694-2&0.57(5)&0.34(3)\\
\hline
\end{tabular}
}
\tablefoot{Numbers in parentheses are one standard deviation in units of the least significant digits.
%\tablefoottext{a}{}
}
\end{table}

\begin{figure*}[!b]
\centering
 \includegraphics [width=0.9\textwidth]{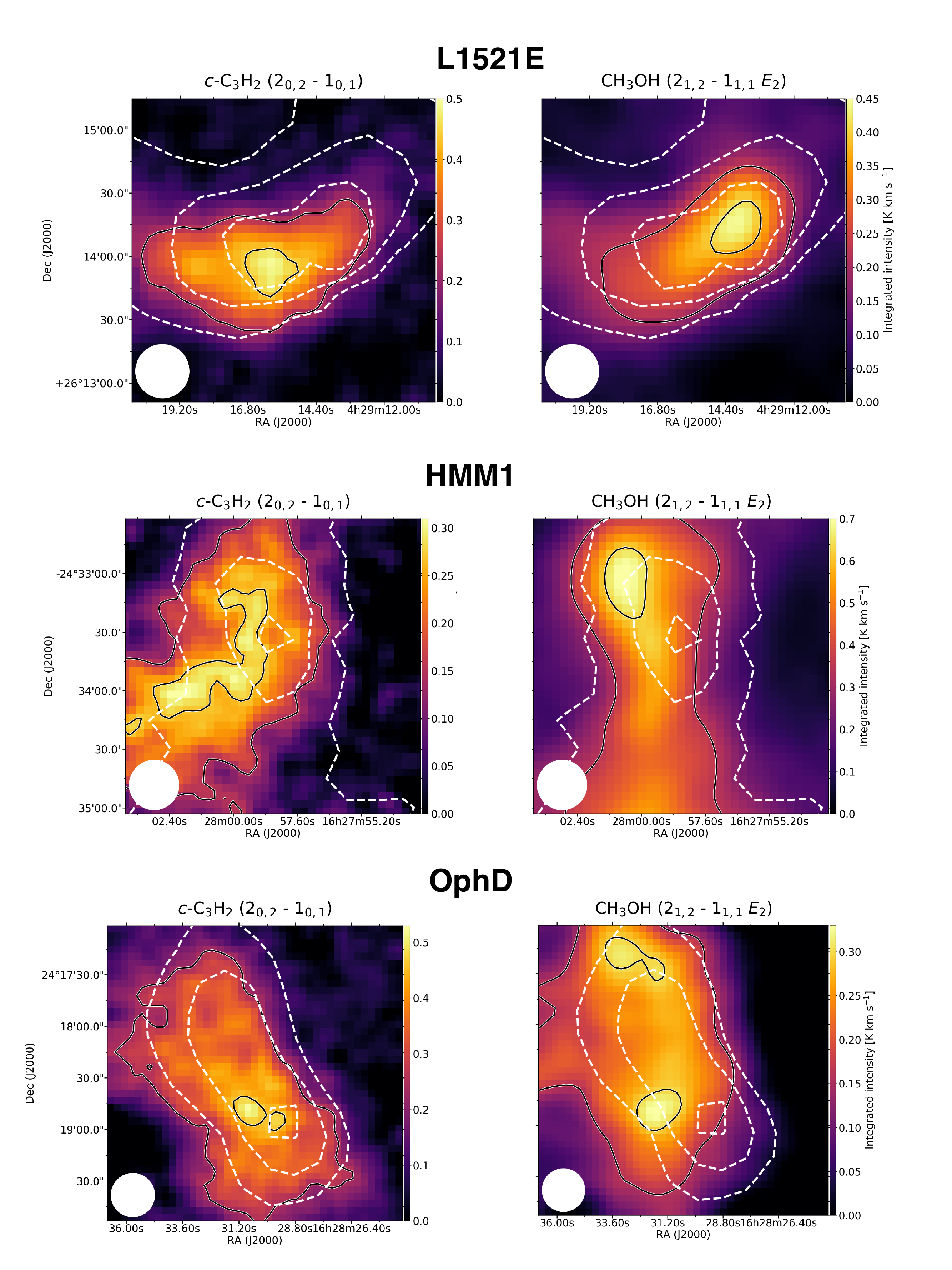}
 \caption{Integrated emission maps of (left) $c$-C$_3$H$_2$ and (right) CH$_3$OH towards the sources in our sample. The solid line contours indicate  90$\%$ and 50$\%$ of the integrated intensity peak for all molecules with the exception of the $2_{0,2} - 1_{0,1}$ (E$_1$) CH$_3$OH transition in L429, where the contours indicate  90$\%$ and 65$\%$ of the integrated intensity peak. The dashed line contours represent the A$_V$ values computed from the {\em Herschel}/SPIRE maps (5, 10, and 15 mag for B68; 20, 40, and 70 mag for L429; 10, 15, and 20 mag for L1521E; 20, 40, and 80 mag for L694-2; 20, 40, and 60 mag for HMM-1; 20, 30, and 40 mag for OphD). The white circle in the bottom left corner of each panel shows the 40$\arcsec$ beam of the SPIRE data.}
  \label{fig:maps}
\end{figure*}

\medskip
\begin{figure*}[htb]\ContinuedFloat
\centering
 \includegraphics [width=0.9\textwidth]{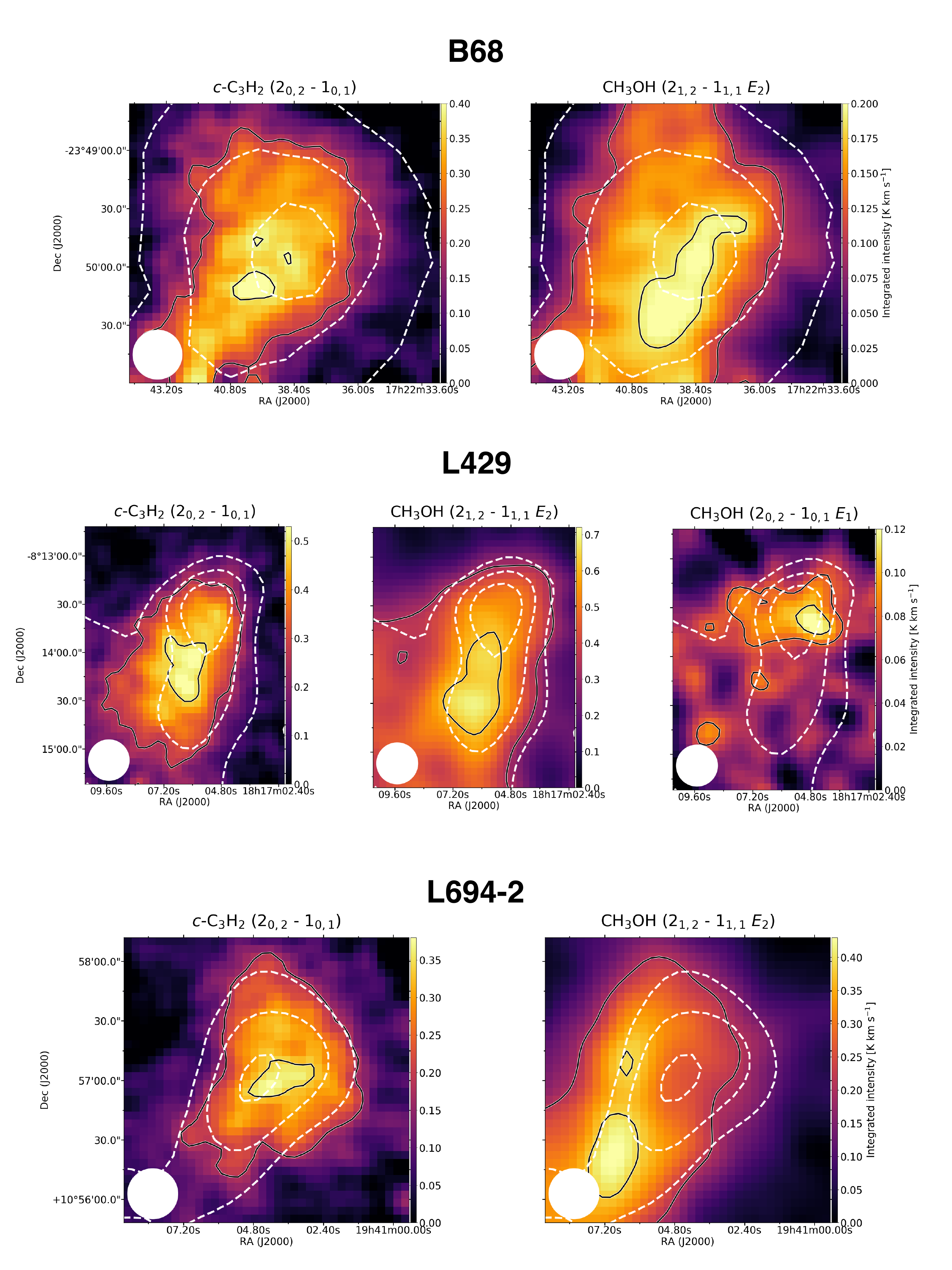}
 \caption{cont. }
\end{figure*}

\begin{figure*}[!b]
\centering
 \includegraphics [width=0.9\textwidth]{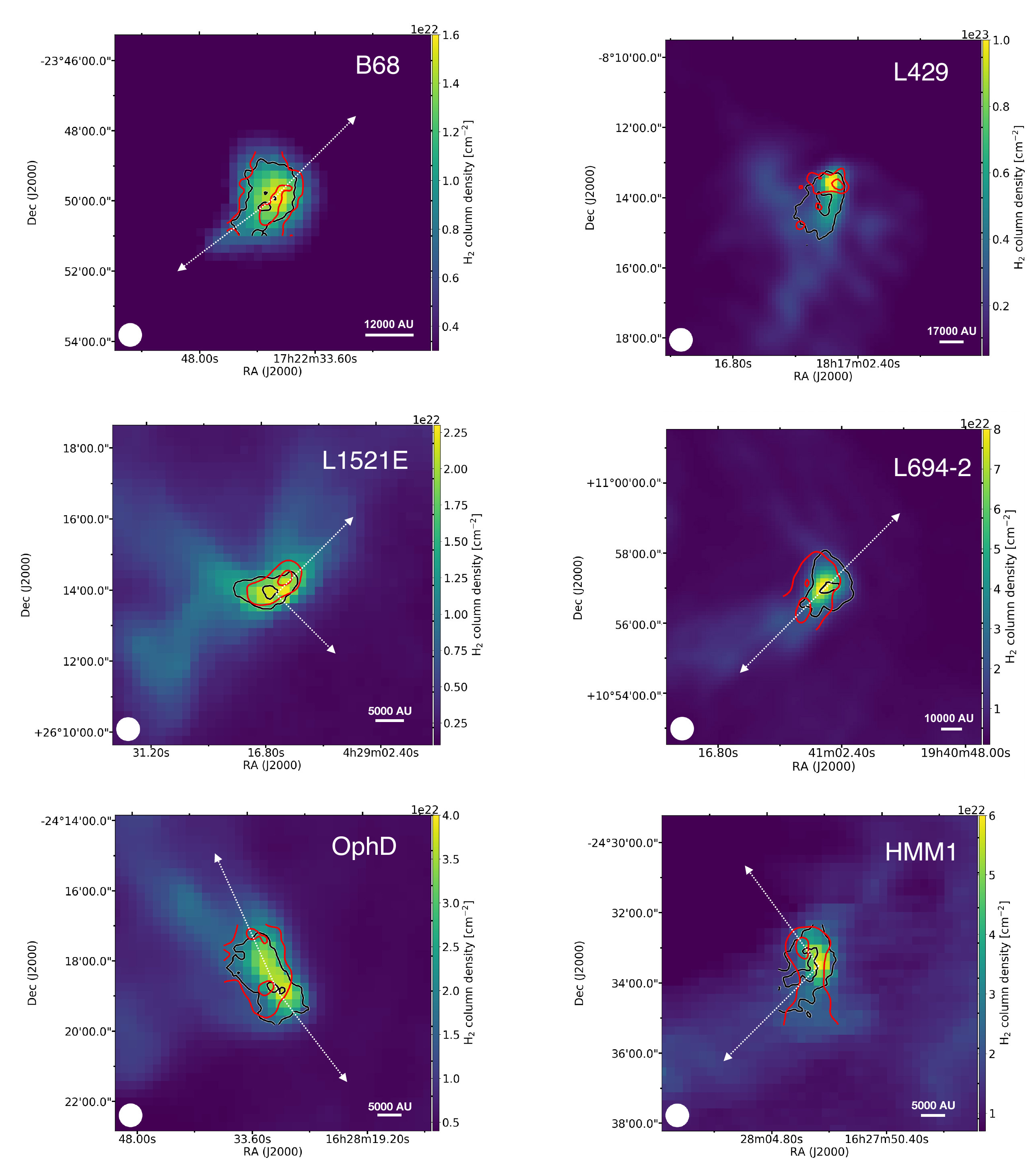}
 \caption{9 $\times$ 9 arcmin$^2$ H$_2$ column density maps computed from {\em Herschel}/SPIRE data. The red and black contours represent   90$\%$ and 50$\%$ of the integrated intensity peak of CH$_3$OH and $c$-C$_3$H$_2$, respectively. The white circle in the bottom left of each panel  shows the 40$\arcsec$ beam of the SPIRE data. The dotted white arrows (not present for L429) indicate the directions in which the abundances profiles discussed in Section 4 were extracted.}
  \label{fig:H2maps}
\end{figure*}

\begin{figure*}[h]
\centering
 \includegraphics [width=1\textwidth]{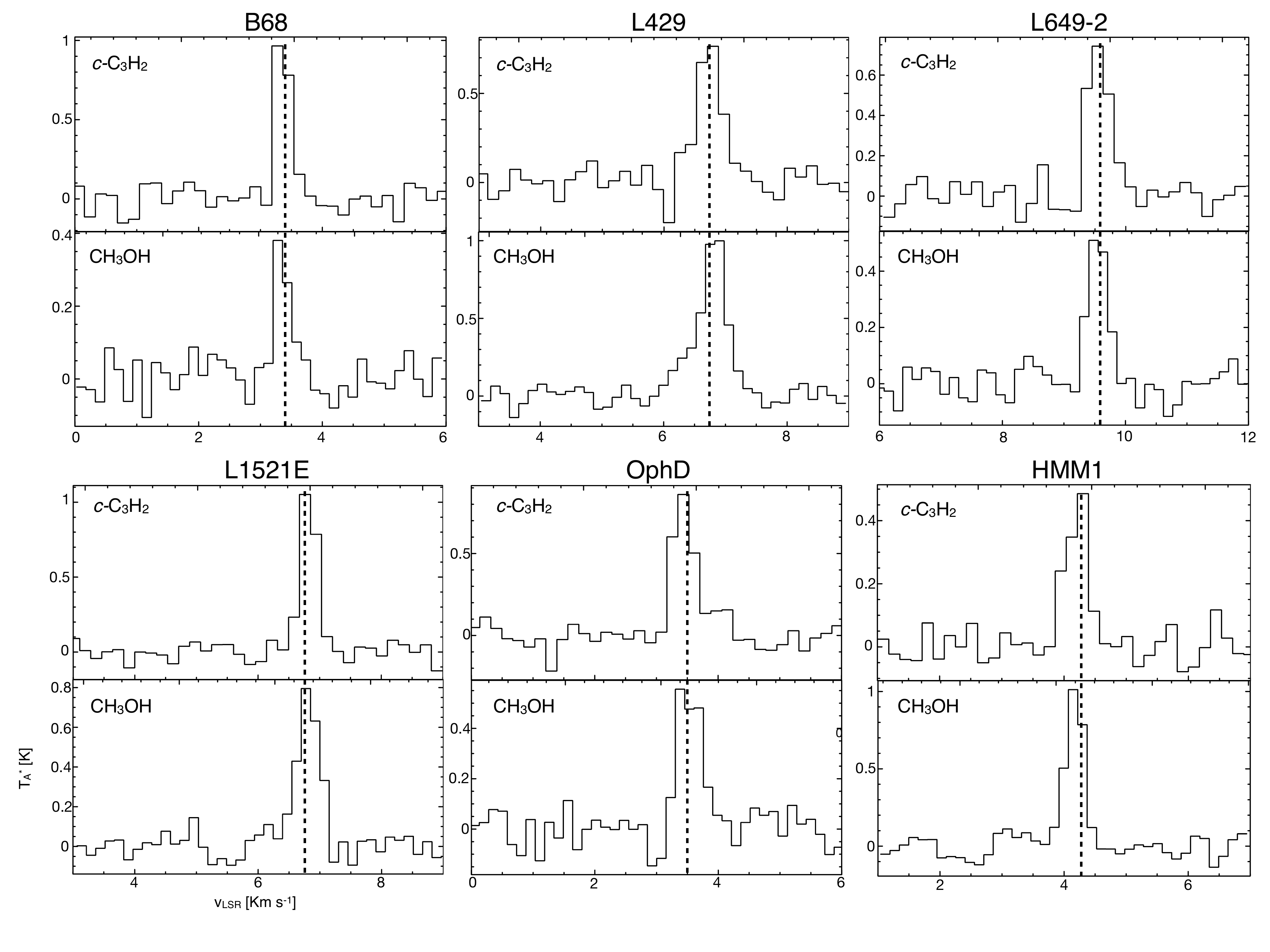}
 \caption{Spectra of the $J_{K_a,K_c}$ = $2_{0,2} - 1_{0,1}$ transition of cyclopropenylidene and of the $J_{K_a,K_c}$ = 2$_{1,2}$-1$_{1,1}$ ($E_2$) transition of methanol towards the dust peak of the sources in our sample. The dotted vertical line shows the systemic velocity of the sources listed in Table~\ref{table:sources}. }
  \label{fig:spectra}
\end{figure*}

\begin{table*}
\caption{Spectroscopic parameters of the observed lines}
\label{table:parameters}
\scalebox{0.95}{
\begin{tabular}{cccccc}
\hline\hline
Molecule & Transition & Rest frequency\tablefootmark{a} & E$_{up}$& A & n$^*$\tablefootmark{b}
\\
&      &(MHz)   & (K)  & ($\times$10$^{-5}$ s$^{-1}$)&cm$^{-3}$\\
\hline
$c$-C$_3$H$_2$  &$J_{K_a,K_c}$ = $2_{0,2} - 1_{0,1}$    &    82093.544(1)   &   6.43     &  1.89&5$\times$10$^5$\\
CH$_3$OH   &  $J_{K_a,K_c}$ = 2$_{1,2}$-1$_{1,1}$ ($E_2$) &   96739.358(2)       &  12.53\tablefootmark{c}   & 0.26&3$\times$10$^4$\\
CH$_3$OH   &  $J_{K_a,K_c}$ = 2$_{0,2}$-1$_{0,1}$ ($E_1$) &   96744.545(2)       &  20.08\tablefootmark{c}   & 0.34&3$\times$10$^4$\\
CCS   &  $N_J$ = 6$_7$-5$_6$ &   81505.17(2)      &  15.39  & 2.43&--\\
C$_4$H   &   $N$ = 9 - 8 ,$J$ = 19/2 - 17/2 ,$F$ = 9 - 9 $\&$ 9 - 8  &   85364.004(3) $\&$ 85364.015(3)       &  20.55 & 0.26&--\\

\hline
\end{tabular}
}
\tablefoot{
\tablefoottext{a}{Frequencies and uncertainties from the CDMS \citep{Mueller2005}}
\tablefoottext{b}{n$^*$ is the critical density, calculated at 10 K.}
\tablefoottext{c}{Energy relative to the ground 0$_{0,0}$, $A$ rotational state.}
}
%$^b$ From \citet{kim05}.\\
\end{table*}

\begin{table}
\caption{Observed line parameters of CH$_3$OH and $c$-C$_3$H$_2$}
\label{table:lines}
\scalebox{0.85}{
\begin{tabular}{c|cccc}
\hline\hline
Molecule & Transition &Area & V$_{LSR}$ & $\Delta$v\\
&   (J'$_{K'_a K'_c}$-J''$_{K''_a K''_c}$)&  (K km s$^{-1}$)&(km s$^{-1}$)&(km s$^{-1}$)\\
\hline
&&\textbf{L1521E}&&\\
$c$-C$_3$H$_2$&$2_{0,2} - 1_{0,1}$&0.40(2)&6.82(1)&0.32(2)\\
CH$_3$OH&2$_{1,2}$ - 1$_{1,1}$&0.35(2)&6.81(1)&0.41(3)\\
&&\textbf{HMM-1}&&\\
$c$-C$_3$H$_2$&$2_{0,2} - 1_{0,1}$&0.23(2)&4.21(2)&0.45(4)\\
CH$_3$OH&2$_{1,2}$ - 1$_{1,1}$&0.39(2)&4.17(1)&0.35(2)\\
&&\textbf{OphD}&&\\
$c$-C$_3$H$_2$&$2_{0,2} - 1_{0,1}$&0.38(3)&3.43(2)&0.41(4)\\
CH$_3$OH&2$_{1,2}$ - 1$_{1,1}$&0.28(2)&3.53(2)&0.45(4)\\
&&\textbf{B68}&&\\
$c$-C$_3$H$_2$&$2_{0,2} - 1_{0,1}$&0.32(2)&3.35(1)&0.22(7)\\
CH$_3$OH&2$_{1,2}$ - 1$_{1,1}$&0.12(2)&3.34(2)&0.27(6)\\
&&\textbf{L429}&&\\
$c$-C$_3$H$_2$&$2_{0,2} - 1_{0,1}$&0.40(3)&6.73(2)&0.47(5)\\
CH$_3$OH&2$_{1,2}$ - 1$_{1,1}$&0.55(3)&6.80(1)&0.52(3)\\
&&\textbf{L694-2}&&\\
$c$-C$_3$H$_2$&$2_{0,2} - 1_{0,1}$&0.35(3)&9.56(2)&0.42(3)\\
CH$_3$OH&2$_{1,2}$ - 1$_{1,1}$&0.23(2)&9.54(2)&0.38(3)\\
\hline
\end{tabular}
}
\tablefoot{Numbers in parentheses are one standard deviation in units of the least significant digit.}
\end{table}

\subsection{L1521E}
The maps of CH$_3$OH and $c$-C$_3$H$_2$ observed towards L1521E are shown in Figure~\ref{fig:maps}. Even if it is not as pronounced as in other cores in our sample, a chemical segregation between the two molecules is present in L1521E, with the methanol peaking in the north-west and the cyclopropenylidene towards the centre or southern part of the core. The spatial distribution of $c$-C$_3$H$_2$ is confirmed to follow those of the maps of C$_4$H  and CCS, and the distribution of the E$_2$ methanol line is consistent with the map of the E$_1$ methanol line shown in the Appendix (Figure~\ref{fig:appendix}). We can hence assume that the moderately high optical depth of the $c$-C$_3$H$_2$ transition reported in Table~\ref{table:tau} is not affecting the morphology of the distribution. As can be seen in the large-scale maps shown in Figure~\ref{fig:H2maps}, the southern part of the core is the most exposed to the ISRF.

\subsection{HMM-1}
Also in HMM-1, methanol and cyclopropenylidene trace different parts of the core, with methanol tracing the eastern part, and the cyclopropenylidene surrounding it, as shown in Figure~\ref{fig:maps}. The emission map of CCS in the Appendix (Figure~\ref{fig:appendix}) confirm that carbon-chain molecules trace the north-west part of HMM-1. The optical depth of the 2$_{1,2}$ - 1$_{1,1}$ (E$_2$) methanol line is moderately high (see Table~\ref{table:tau}). In order to check whether the optical depth has an impact on the structure of its emission, we present the map of the $2_{0,2} - 1_{0,1}$ (E$_1$) transition of methanol in Figure~\ref{fig:appendix}, which shows the same spatial distribution of the 2$_{1,2}$ - 1$_{1,1}$ (E$_2$) line.
However, the large-scale maps in Figure~\ref{fig:H2maps}, seem to contradict what has been observed towards L1544. The density in HMM-1 has a sharper drop towards the eastern side where methanol peaks. This can however be explained by the presence of the B-type stars $\rho$ Oph and HD 147889 towards the west with respect to HMM-1 \citep{deGeus92,Preibisch08}. The increase in  the kinetic temperature towards HD 147889 in the $\rho$ Ophiucus molecular cloud was measured by observing hyperfine anomalies in the OH 18 cm line \citep{ebisawa15}. Furthermore, the effect of HD 147889 on the region was studied with maps of CI, CII, and CO, showing an arrangement that is in contrast to the prediction of plane-parallel PDR models and can be explained as significant density gradients, as well as different chemical evolution \citep{kamegai03}. When considering that HMM-1 is embedded in a non-homogeneous ISRF because of a strong illumination from the west due to the stars,  for HMM-1 the differentiation between CH$_3$OH and $c$-C$_3$H$_2$ also follows the trend seen in L1544 and in the other cores described in this paper.
The asymmetry of the methanol distribution in HMM-1 is discussed in \cite{harju20}.

\subsection{OphD}
The chemical segregation between $c$-C$_3$H$_2$ and CH$_3$OH in OphD is clearly visible in Figure~\ref{fig:maps}. The methanol emission is elongated and has two emission peaks, while that of cyclopropenylidene is more concentrated towards the dust peak of the core. Figure~\ref{fig:appendix}  shows the emission maps of C$_4$H and CCS, which  confirm that carbon-chain molecules in OphD trace the south-western part of the core. The emission of the E$_1$ methanol line could not be mapped because it is too weak. In the large-scale map in Figure~\ref{fig:H2maps} we show  that the methanol emission is farther away from the edge of the core, hence the more illuminated side where the carbon chain molecules peak.
As already discussed for HMM-1, the presence of the B-type stars $\rho$ Oph and HD 147889 also has an effect on OphD (Figure~\ref{fig:H2maps}) with the difference that in OphD the sharper drop in density, hence the more illuminated side, faces the west where  the B-type stars are also located, and both contribute to an active hydrocarbon chemistry.

\subsection{B68}
Figure~\ref{fig:maps} shows the emission maps of $c$-C$_3$H$_2$ and CH$_3$OH, observed towards the inner 80 arcsec $\times$ 80 arcsec of B68 with the IRAM 30m telescope. The distribution of the two molecules is similar and without a preferential position around the starless core. The spatial distribution of $c$-C$_3$H$_2$ is confirmed by the map of C$_4$H shown in   Figure~\ref{fig:appendix}. The emission of the E$_1$ methanol line could not be mapped because it is too weak. The large-scale map in Figure~\ref{fig:H2maps} confirms that B68 is indeed an isolated globule and not part of a more complex density structure, while the other cores
in our sample are.  As a consequence, we can infer that a homogeneous ISRF surrounds B68.

\subsection{L429}
No chemical segregation can be inferred from the observed emission maps of $c$-C$_3$H$_2$ and CH$_3$OH $2_{1,2} - 1_{1,1}$ (E$_2$) towards L429 as both molecules seem to trace the same region within the core, as shown in Figure~\ref{fig:maps}.

The central H$_2$ column density of L429 is seven times higher than B68, the least dense core in our sample, and about a factor of 2 higher than all the other cores, with the exception of L694-2 and L1544. Given the high density of the core, and the high brightness temperature of the lines mapped, the absence of chemical differentiation could be an effect due to the optical depth of the lines, or self-absorption. As was done for HMM-1, also in this case we   checked the map of the $2_{0,2} - 1_{0,1}$ (E$_1$) transition of methanol, shown in Figure~\ref{fig:maps}. The $2_{0,2} - 1_{0,1}$ (E$_1$) transition of methanol shows a different spatial distribution with respect to the $2_{1,2} - 1_{1,1}$ (E$_2$) transition, hinting at the fact that the spatial distribution of the $2_{1,2} - 1_{1,1}$ (E$_2$) transition might be affected by optical depth and/or self-absorption. We  investigated this issue further and used the CASSIS software\footnote{http://cassis.irap.omp.eu} \citep{vastel15} to model the observed spectra of methanol towards the dust peak in L429. It is not possible to reproduce the relative intensity among the three methanol lines ($2_{0,2} - 1_{0,1}$ E$_1$,  $2_{1,2} - 1_{1,1}$ E$_2$, and $2_{0,2} - 1_{0,1}$ A$^+$) at 96.7 GHz without using a non-LTE approach. When modelling the spectra with the LTE assumption with an excitation temperature of either 5 or 4 K, we overestimate the intensity of the $2_{0,2} - 1_{0,1}$ (E$_1$) line, the highest in energy among the three. This is not the case towards all the other sources in our sample, where the intensity ratio among the three methanol lines can always be reproduced with LTE, assuming an excitation temperature of 5 K. 
When using the non-LTE radiative transfer code RADEX \citep{radex} within the software CASSIS, we can reproduce well the spectrum assuming a T$_{kin}$ of 10 K, a H$_2$ volume density of 6$\times$10$^5$ cm$^{-3}$ \citep{Crapsi05}, and a N(CH$_3$OH) of $\sim$5$\times$10$^{13}$ cm$^{-2}$. The resulting T$_{ex}$ is 9.5 K for the $2_{1,2} - 1_{1,1}$ (E$_2$) transition, and 8 K for the $2_{0,2} - 1_{0,1}$ (E$_1$) transition. Furthermore, the resulting optical depth for the $2_{1,2} - 1_{1,1}$ (E$_2$) line is only 0.3, suggesting that the line is not necessarily optically thick, but rather that the system is  out of LTE.\\
The presence of multiple components along the line of sight might also play a role towards L429.  \cite{Redaelli18}  show that the line width of N$_2$H$^+$ in L429 is broader than in L183 and L694-2. This could be due to environmental effects (e.g. higher levels of turbulence), but also to the presence of multiple components along the line of sight. 
We note that the double peak seen in N$_2$H$^+$ (1-0) has been interpreted as due to self-absorption in a contracting cloud, based on detailed non-LTE modelling \citep{Keto08, Keto10, Keto15}. Nevertheless, \cite{Crapsi05} showed that the broad lines of N$_2$H$^+$ towards the dust peak of L429 have a hint of a double peak while not showing the classical blue-asymmetry attributed to infall, thus suggesting the presence of two velocity components. Considering   that the H$_2$ column density of L429 is exceptionally high within our sample, as well as the results of \cite{Crapsi05} and \cite{Redaelli18}, it is plausible to assume the presence of multiple components along the line of sight.

In Figure~\ref{fig:appendix} we show the emission maps of CCS and C$_4$H, which both trace the same region traced by $c$-C$_3$H$_2$. 
Figure~\ref{fig:H2maps} shows the large-scale maps of N(H$_2$) and T$_{dust}$ computed from SPIRE with the contours of $c$-C$_3$H$_2$ and CH$_3$OH observed with the 30m telescope. Because of the  above-mentioned     optical depth effects, we chose to plot the contours of the $2_{0,2} - 1_{0,1}$ (E$_1$) transition of methanol instead of the $2_{1,2} - 1_{1,1}$ (E$_2$). 
Contrary to what happens in L1544, in L429 methanol  traces the part of the core that is facing the steepest decrease in H$_2$ column density, the north side, and hence the most illuminated part according to the SPIRE maps. Cyclopropenylidene instead traces the southern part of L429. The discrepancy among the two sources might be due to the different environment surrounding the two cores. The Aquila Rift is an active star-forming region that does not allow us to assume a homogeneous ISRF around L429. The closest star to L429, IRAS 18158-0813, is less than 0.5$^\circ$ towards the east \citep{kwok97}, and the brightest nearby source  is a C-star, IRAS 18156-0653, located at 1.5$^\circ$ towards the north \citep{kim13}. On the contrary, L1544 does not have any protostars or young stars nearby, and the asymmetry of the $c$-C$_3$H$_2$--CH$_3$OH distribution correlates well with the density structure that surrounds the core \citep{Spezzano16}.

\subsection{L694-2}
The chemical differentiation between methanol and cyclopropenylidene in L694-2 is clearly shown in Figure~\ref{fig:maps}. While methanol traces the south-eastern side of the core, cyclopropenylidene traces  the central and western parts of the core. The spatial distribution of $c$-C$_3$H$_2$ is confirmed by the map of CCS shown in  Figure~\ref{fig:appendix}. The emission of the E$_1$ methanol line could not be mapped because it is too weak. When looking at the large-scale maps presented in Figure~\ref{fig:H2maps}, it is clear that the segregation in L694-2 is similar to that observed in L1544 \citep{Spezzano16}, with the methanol tracing a side of the core that is more shielded from the ISRF with respect to the side traced by cyclopropenylidene.

\section{Abundances profiles} 
\begin{figure*}[h]
\centering
 \includegraphics [width=0.9\textwidth]{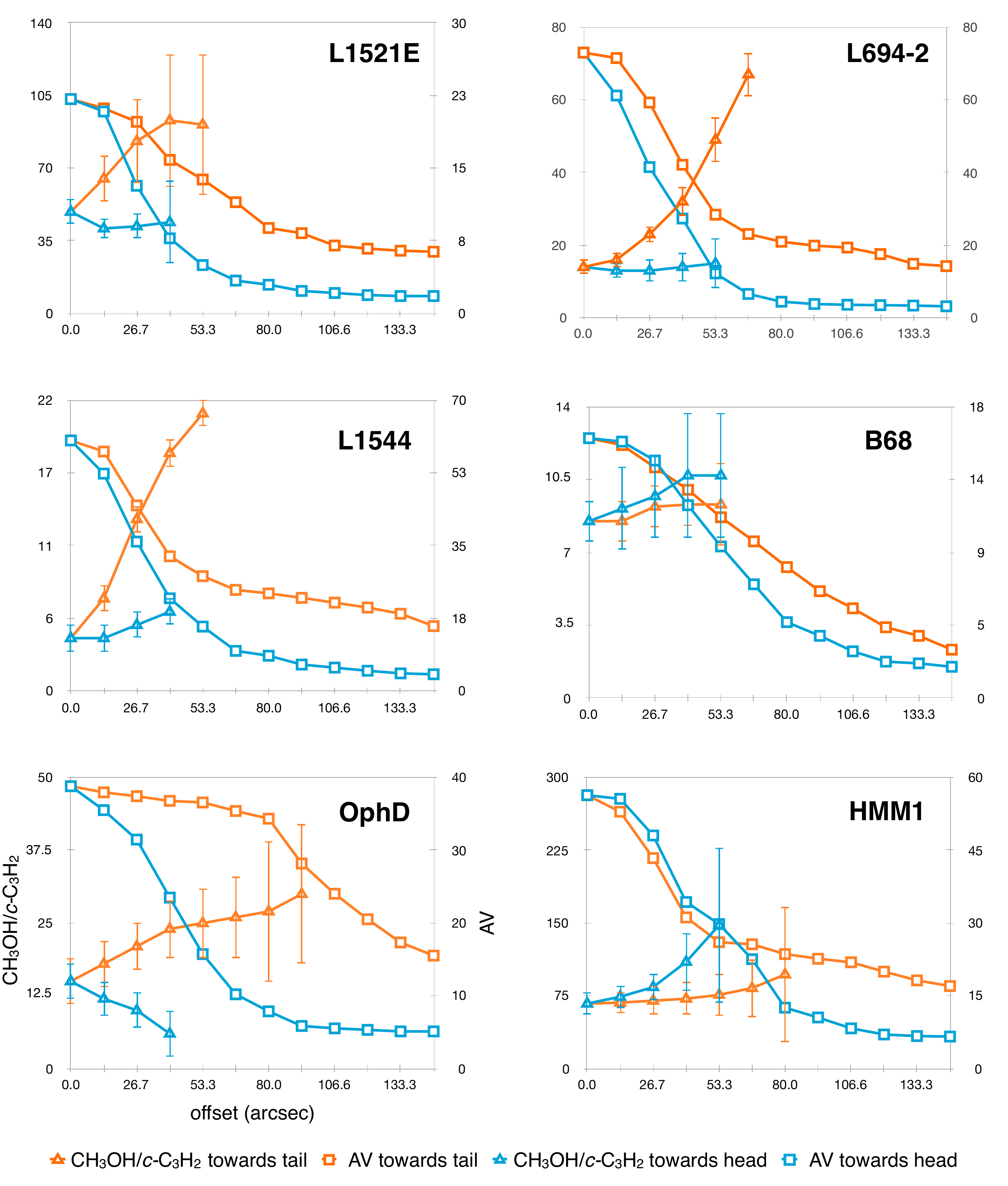}
 \caption{N(CH$_3$OH)/N($c$-C$_3$H$_2$) abundance ratios and the visual extinction A$_V$ extracted towards both the tail and the head in all the cores in our sample. The dotted white arrows in Figure~\ref{fig:H2maps} indicate the directions in which the abundances profiles were extracted. }
  \label{fig:profiles}
\end{figure*}

In order to study quantitatively the dependence of the segregation between $c$-C$_3$H$_2$ and CH$_3$OH on the large-scale structure around the core, we  computed the column density maps for both molecules in each core in our sample, and in L1544. Subsequently, we   extracted the  N(CH$_3$OH)/N($c$-C$_3$H$_2$) ratios in two directions, starting from the core's dust peak. These directions were defined following the cometary shape of starless and pre-stellar cores, already recognised in \cite{Crapsi05}, and are indicated in Figure~\ref{fig:H2maps} by dotted white arrows. They either point towards the `head' or the `tail' of the comet-like structure of the core. The head is intended as the sharp drop in H$_2$ column density, while the tail is a shallower drop. In the case of B68 there is not a proper tail because the core is isolated, and the profile  defined as tail in Figure~\ref{fig:profiles} is the one pointing towards the south-east in Figure~\ref{fig:H2maps}. Given the small extent of the methanol $2_{0,2} - 1_{0,1}$ (E$_1$) line map, it was not possible to extract the profiles for L429. For L1544, the data used in this section were extracted in the profiles shown in Figure 1 of \cite{Spezzano16}, with the head being the profile pointing towards the south-west, and the tail being the profile pointing towards the north-east.

The column densities were  calculated following the formula reported in \cite{Mangum15} for the non-optically thin approximation.
The total column density is defined as

\begin{equation}
%\tau &=& \ln\left(\frac{J(T_{ex}) -  J(T_{bg})}{J(T_{ex}) - J(T_{bg}) - T_{mb}}\right)\\
N_{tot} = \frac{8\pi\nu^3Q_{rot}(T_{ex})W}{c^3A_{ul}g_u}\frac{e^{\frac{E_u}{kT}}}{J(T_{ex}) -  J(T_{bg})}\frac{\tau}{1-e^{-\tau}},\\
\end{equation}

\noindent
where $\tau$ is the optical depth of the line defined in \ref{eq:1}, $c$ is the speed
of light, $A_{ul}$  is the Einstein
coefficient of the transition, $g_u$ is the degeneracy of the upper state, $E_u$ is the
energy of the upper state, and $Q_{rot}$ is the partition function of the molecule at the given temperature $T_{ex}$. To calculate $N_{tot}$ as we did for $\tau$, we assumed a $T_{ex}$ of 5 K for both molecules. The effect of using different excitation temperatures for methanol and cyclopropenylidene on the derived column density ratios was tested and found to be not significant (when decreasing the temperature by 1 K, the effect on the ratio was of the order of 20$\%$). The error introduced by using a constant excitation temperature to calculate the column density map in a pre-stellar core was found to be  negligible in a previous study of L1544 (see   the Appendix of \citealt{Redaelli19}). By using
these expressions, we assumed that the source fills the beam, and the emission obeys LTE. The values of methanol and cyclopropenylidene column densities, as well as the values of A$_V$, are averaged within the SPIRE data beam (40\arcsec) and are extracted with a step of 1/3 of a beam. Unfortunately, we cannot use non-LTE radiative transfer codes to derive the column densities of CH$_3$OH and $c$-C$_3$H$_2$ towards the different profiles in our sources because the physical models available are spherically symmetric (usually a Bonnor-Ebert sphere), and hence cannot reproduce the asymmetry in our maps. In order to assess the impact that non-LTE effects might have on our results, we  calculated the deviation of the LTE assumption in a range of volume densities between 1$\times$10$^4$ and 1$\times$10$^6$ cm$^{-3}$. We find that while the LTE assumption might underestimate the N(CH$_3$OH)/N($c$-C$_3$H$_2$) abundance ratio by a factor of 2-3, it does not vary strongly with volume density, and hence it does not have an effect on the trends that we  discuss in this section and show in Figure~\ref{fig:profiles}. More details are given in  Appendix~\ref{prova}.

\begin{figure}[h]
\centering
 \includegraphics [width=0.5\textwidth]{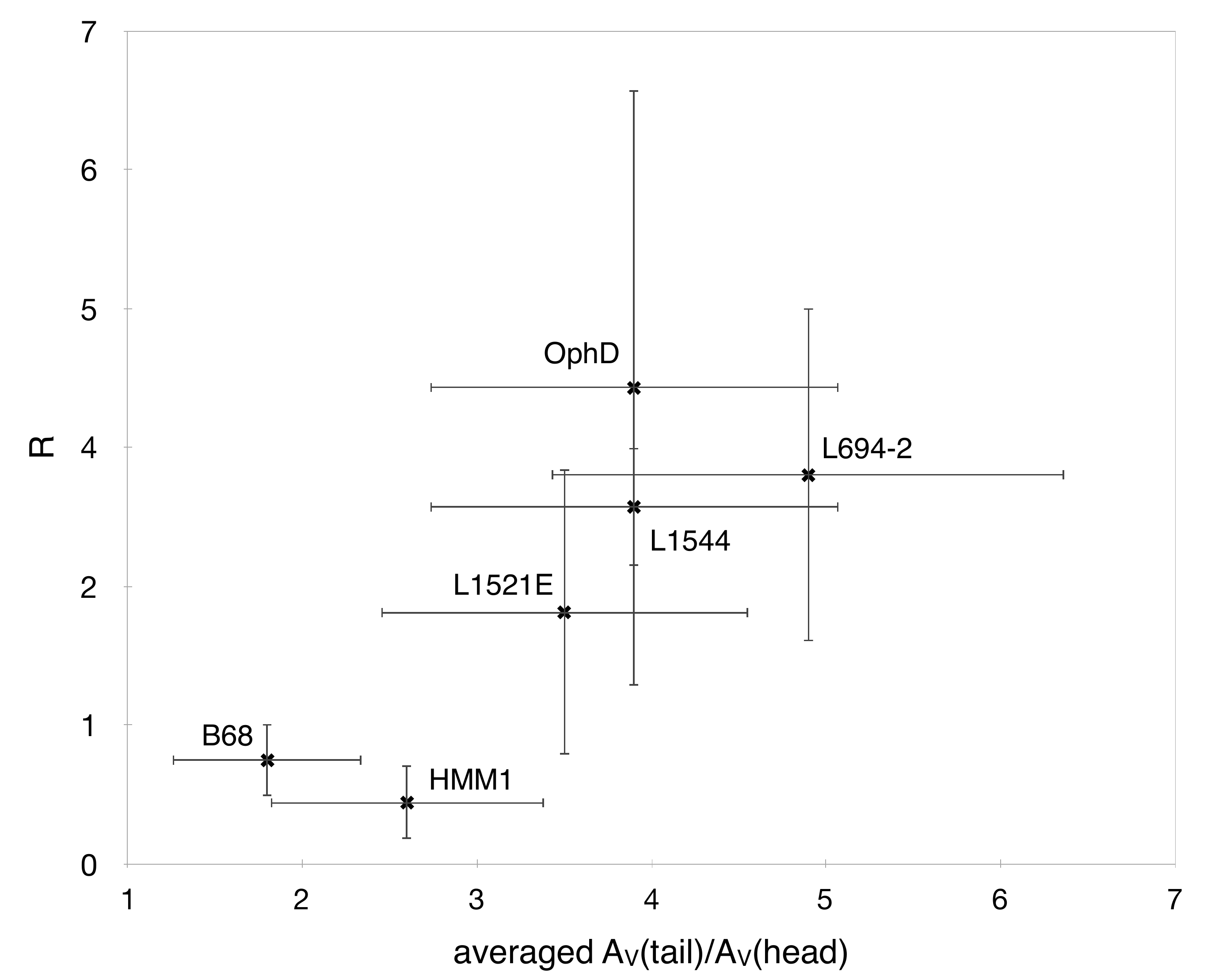}
 \caption{Ratio of the abundance ratios N(CH$_3$OH)/N($c$-C$_3$H$_2$) in the head and tail profiles at the largest observed offset (R) plotted against the ratio of the averaged A$_V$ towards the head and tail of each core.}
  \label{fig:ratio}
\end{figure}

Figure~\ref{fig:profiles} shows the profiles of the N(CH$_3$OH)/N($c$-C$_3$H$_2$) abundance ratios as well as the visual extinction A$_V$ towards both the tail and the head in all the cores in our sample. The errors on the A$_V$ are comparable to the errors estimated for the H$_2$ column density computed from the SPIRE data, namely $\sim$20\% (see e.g. \citealt{Chantzos18}). The error bars on the A$_V$ values are not shown to avoid overcrowding the plots.
In all cores where a substantial difference in the N(CH$_3$OH)/N($c$-C$_3$H$_2$) abundance ratio between the tail and the head is present (L1521E, L694-2, L1544, and OphD), a correspondent substantial difference can be seen in the A$_V$ at larger radii. B68, being an isolated core, does not have a tail or head structure, and does not have a differentiation among the two molecules. HMM-1 shows a hint of anti-segregation, with the N(CH$_3$OH)/N($c$-C$_3$H$_2$) abundance ratio increasing towards the head rather than towards the tail. As already mentioned, this can be explained by the active environment where HMM-1 is embedded, where the presence of two  B-type stars towards the west makes  the illumination on that side of the core more prominent, despite its density structure, and hence does   describe the segregation based solely on the head and tail structure of the core. In the case of OphD, also located  in Ophiucus nearby HMM-1, the head is facing the west, and hence the illumination of the nearby stars contributes to the spatial segregation among methanol and cyclopropenylidene in the core.
This effect can be seen in Figure~\ref{fig:ratio}, where we plot for each source the ratio of N(CH$_3$OH)/N($c$-C$_3$H$_2$) in the two profiles (tail and head) at the largest observed offset in the 30m data (R), against the ratio among the A$_V$ towards the tail and the head of the core. The three cores embedded in a homogeneous ISRF (L1521E, L1544, and L694-2) show a clear increase in R with increasing A$_V$ ratio in the tail. OphD shows the largest R in our sample even without having the highest A$_V$ ratio in the tail, hinting at the presence of an additional factor that increases the segregation between CH$_3$OH and $c$-C$_3$H$_2$. In our opinion this effect is due to the presence of the stars nearby. B68 shows no segregation (R=1), and confirms the linear correlation among R and the A$_V$ ratio shown by L1521E, L1544, and L694-2.

\section{Methanol in starless cores}\label{results2}
Non-thermal desorption mechanisms play a crucial role in the development of chemical complexity in starless cores, and starless cores set the initial conditions for the chemical complexity in the processes of star and planet formation. It is therefore critical to understand the interplay of the different chemical and physical processes at play in starless cores. Methanol has been observed in several starless cores \citep{Soma15,Vastel14}, and mapped in a few \citep{Bizzocchi14, tafalla06}. With the present dataset we can correlate the distribution of methanol around six starless cores, seven if we also include  L1544 in the sample, to the physical properties of the cores, with the aim of putting some observational constraints on the chemical models.
In Table~\ref{table:properties} we report the visual extinction, the T$_{dust}$ and the N(CH$_3$OH) values at the methanol peak, as well as the distance of the methanol peak from the {\em Herschel} dust peak, the central H$_2$ column density for each core, and the methanol abundance at the methanol peak. The A$_V$ at the methanol peak varies from 14 in B68 to 96 in L429 (at the peak of the $2_{0,2} - 1_{0,1}$ (E$_1$) transition); this large spread suggests that the processes responsible for the desorption of methanol in the gas phase are not too sensitive to H$_2$ column density variations within the column density ranges in Table~\ref{table:properties}. 
This result is consistent with the expectation that CH$_3$OH requires high enough A$_V$ for efficient surface production and release in the gas phase. \cite{Vasyunin17} presented the results of a gas-grain astrochemical model which includes the reactive desorption to model the abundance of complex organic molecules towards pre-stellar cores. The results presented in \cite{Vasyunin17} show that by using the physical structure for the pre-stellar core L1544 \citep{Keto10} complex organic molecules peak at radial distances between 0.01 and 0.03 pc, which in L1544 correspond to a visual extinction of 7 magnitudes. We note that the extinction given in Figure 2 of \cite{Vasyunin17} is the model extinction seen by the core, and does not consider the cloud where the core is embedded. With the {\em Herschel} observations, instead, we see the total extinction. The model extinction from \cite{Vasyunin17} needs to be multiplied by a factor of two to be compared with the value observed with {\em Herschel}, and reported in Table~\ref{table:properties} of this paper.

\begin{figure}[h]
\centering
 \includegraphics [width=0.5\textwidth]{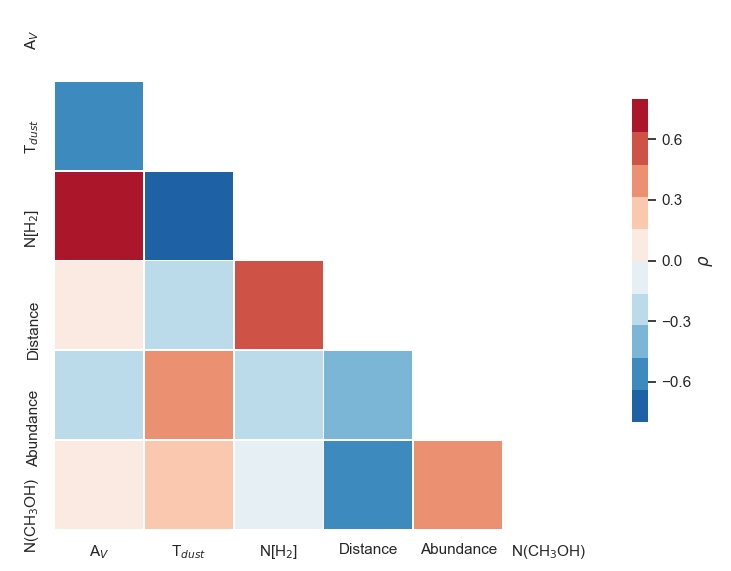}
 \caption{Diagonal correlation matrix showing the Spearman correlation coefficients $\rho$ among the methanol peak properties in Table~\ref{table:properties}.}
  \label{fig:correlation}
\end{figure}

To check if there is any interdependence among the parameters listed in Table~\ref{table:properties}, we  calculated the Spearman correlation coefficient $\rho$ among them. The results are shown in a diagonal correlation matrix in Figure~\ref{fig:correlation}.
The only strong correlations found are between the A$_V$ at the methanol peak and the central H$_2$ column density, and the T$_{dust}$ at the methanol peak and the central H$_2$ column density, both $\rho$ $\simeq$ 0.7. Given the lack of correlation between the abundance of methanol and the central H$_2$ column density of the core, the correlations between the A$_V$ and T$_{dust}$ at the methanol peak and the central H$_2$ column density are very likely to be dependent on the evolutionary state of the core rather than on the desorption mechanism of methanol.
A moderate correlation ($\rho$ $\simeq$ 0.5) is found between the central density and the distance of the methanol peak from the dust peak, which suggests that the location of the methanol peak, and hence the formation of methanol in starless and pre-stellar cores, depends on the physical structure and evolution of the core. Interestingly, the methanol abundance with respect to molecular hydrogen does not show any correlation with the other parameters listed in Table~\ref{table:properties}. Additional observations are needed to put more constraints on the interdependence of the physical and chemical evolution in the earliest stages of star formation. In particular large maps of different isotopologues of carbon monoxide will be useful to understand the link between the formation--desorption of methanol and the CO freeze-out.

\begin{table*}
\centering
\caption{Methanol peak properties }
\label{table:properties}
\scalebox{0.95}{
\begin{tabular}{c|cccccc}
\hline\hline
&A$_V$\tablefootmark{a}&T$_{dust}$\tablefootmark{b}& Central N(H$_2$)\tablefootmark{c}&N(CH$_3$OH)\tablefootmark{d}& Distance from dust peak &  N(CH$_3$OH)/ N(H$_2$)\tablefootmark{e} \\
&[mag]&[K]&[10$^{22}$cm$^{-2}$] &$\times$10$^{13}$ cm$^{-2}$&[au]&$\times$10$^{-9}$\\
\hline \hline
L1521E & 23(3) &12(2)& 2.5(4) &12.0(7)& 2000&6(1)\\
HMM-1 &39(6) &12(2)& 6.0(9)&18(2) & 6000 & 5(1)\\
OphD & 26(4) &12(4)&  4.0(6) &8.0(6) &7000 &3.3(8) \\
B68 & 14(2)&13(2)&1.4(2) &5(1)& 6000&4(1)\\
L429 & 96(14) &10(1)& 10(1) &2.8(7)&7000&0.3(1)\\
L694-2 & 23(3) &11(2)& 8(1)&11(1)& 13000&5(1)\\
L1544 & 36(5) &11(2)& 6.0(9) &12(1)& 4000 & 3.5(9)\\
\hline
\end{tabular}
}
\tablefoot{ Numbers in parentheses are one standard deviation in units of the least significant digits.
\tablefoottext{a}{A$_V$ at methanol peak.}
\tablefoottext{b}{T$_{dust}$ at methanol peak.}
\tablefoottext{c}{Value computed from {\em Herschel}/SPIRE observations towards the dust peak.}
\tablefoottext{d}{N(CH$_3$OH) at methanol peak.}
\tablefoottext{e}{The methanol abundance at the methanol peak is averaged within the {\em Herschel} beam (40$\arcsec$).}
}
\end{table*}

\section{Conclusions}\label{conclusions}
We compared the spatial distribution of methanol and cyclopropenylidene in two starless and four pre-stellar cores at different evolutionary stages and embedded in different environments, and conclude that large-scale effects have a direct impact on the chemical segregation that we observe at core-scale.
Methanol and cyclopropenylidene trace different regions and give us complementary views of dense cloud cores, thus enriching our understanding of the complex interplay between physics and chemistry in the early stages of star formation. The conclusions that we draw by observing crucial molecular tracers prove once again the importance of astrochemistry in the field of star formation. With the present study we showed that there is a correlation among the abundance ratio of methanol and cyclopropenylidene within the core and the large-scale density structure around the core. More observational studies are needed in order to better constrain  the physical structure of these core, and in particular how they merge into the parent cloud. This is needed for the radiative transfer modelling of the molecular emission maps. 
\newline
We  also used our dataset to put observational constraints on the non-thermal desorption of methanol in starless cores. We find that the methanol peaks in a wide range of H$_2$ column densities in the different cores in our sample, suggesting that while the formation of methanol requires a high visual extinction (A$_V$ $>$7 mag) for both its production on the grains and its release in the gas-phase, the H$_2$ column density is not a crucial parameter in the non-thermal desorption of methanol in starless cores.
Our results suggest instead that the central density of the core might have an influence on the distance between the dust peak and the methanol peak. Given the importance of methanol in the development of complex organic chemistry in star-forming regions, more observational studies, both with single-dish telescopes and interferometers, should be performed in order to put more constraints on the chemical models, and finally shine some light on the release in the gas-phase of methanol in starless cores.

\begin{acknowledgements}
The authors wish to thank the anonymous referee and the editor for insightful comments.
\end{acknowledgements}

{}
\newpage

\begin{appendix}

\section{Additional emission maps}
We present here additional maps   observed with the IRAM 30m telescope towards the six starless cores in our sample (see Figure~\ref{fig:appendix}). The properties of the observed transitions are listed in Table~\ref{table:properties}.

\begin{figure*}[!b]
\centering
 \includegraphics [width=0.9\textwidth]{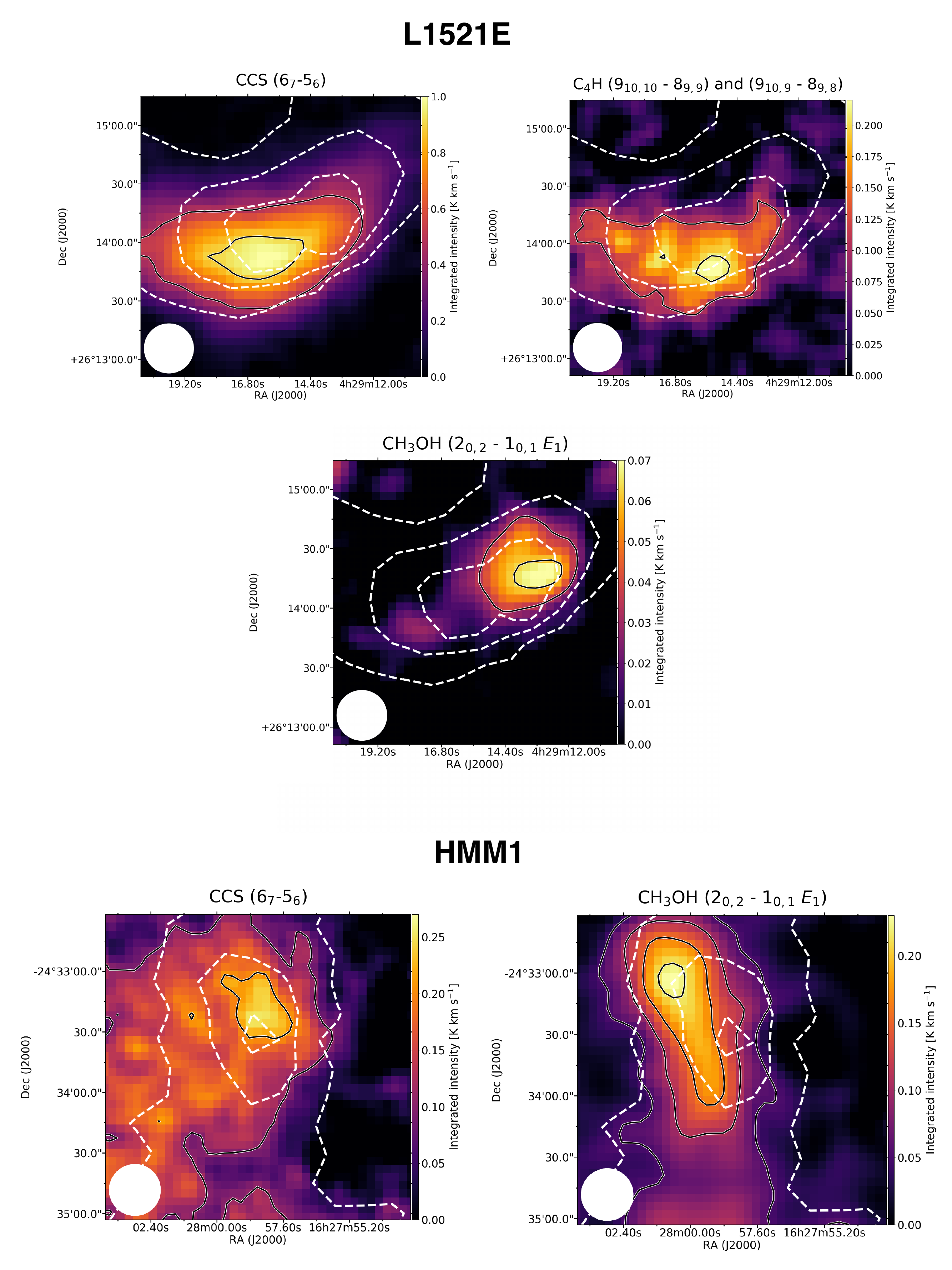}
 \caption{Integrated emission maps of CCS and C$_4$H towards the sources in our sample. The solid line contours indicate   90$\%$ and 50$\%$ of the integrated intensity peak with the exception of the C$_4$H transition in L429, where the contours indicate   90$\%$ and 70$\%$ of the integrated intensity peak. The dashed line contours represent the A$_V$ values computed from the {\em Herschel}/SPIRE maps (5, 10, and 15 mag for B68; 20, 40, and 70 mag for L429; 10, 15, and 20 mag for L1521E; 20, 40, and 80 mag for L694-2; 20, 40, and 60 mag for HMM-1; 20, 30, and 40 mag for OphD). The white circle in the bottom left of each panel   shows the 40$\arcsec$ beam of the SPIRE data.}
  \label{fig:appendix}
\end{figure*}

\medskip
\begin{figure*}[htb]\ContinuedFloat
\centering
 \includegraphics [width=0.9\textwidth]{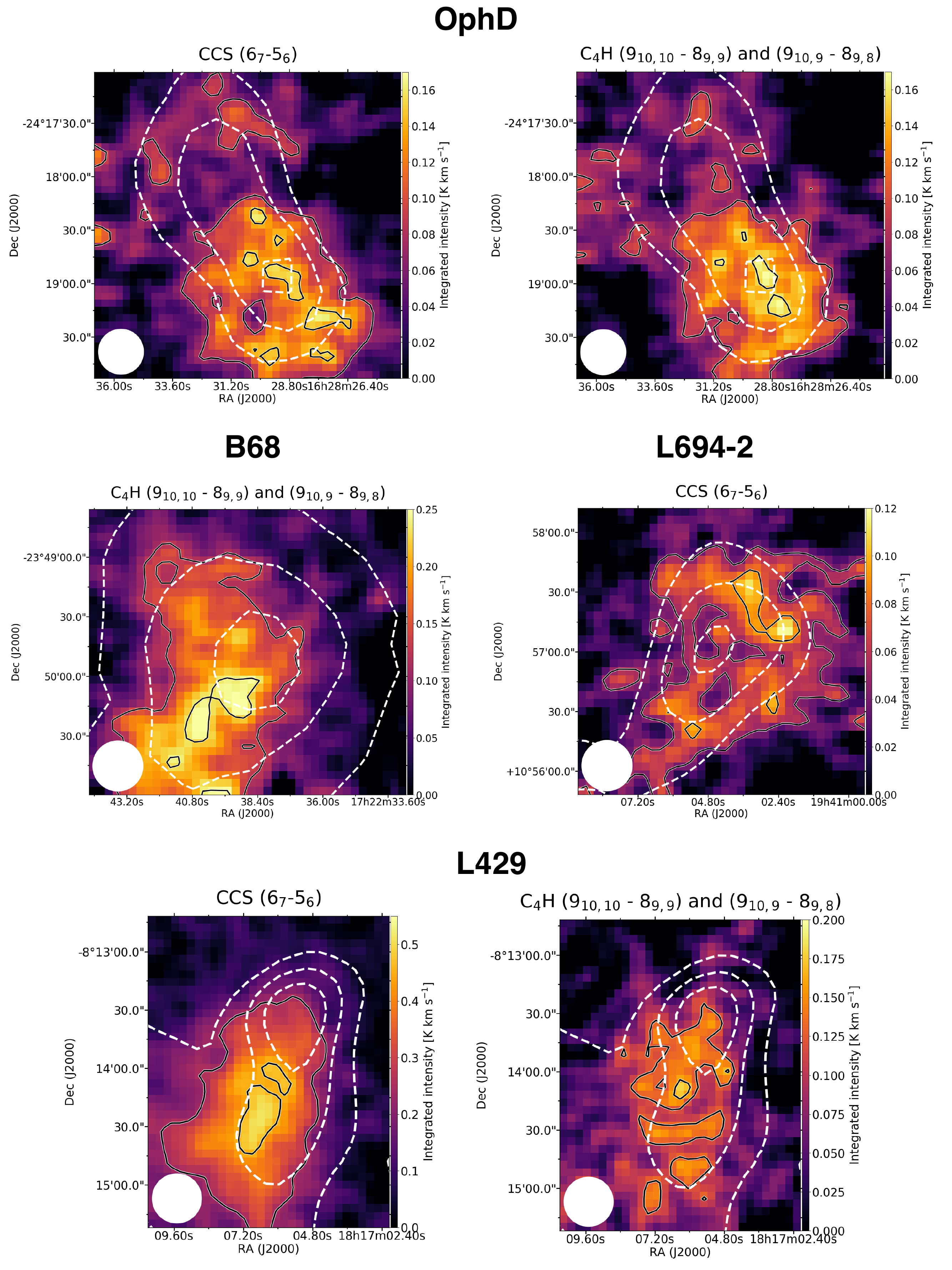}
 \caption{cont. }
\end{figure*}

\section{T$_{dust}$ maps}
\begin{figure*}[h]
\centering
 \includegraphics [width=1.0\textwidth]{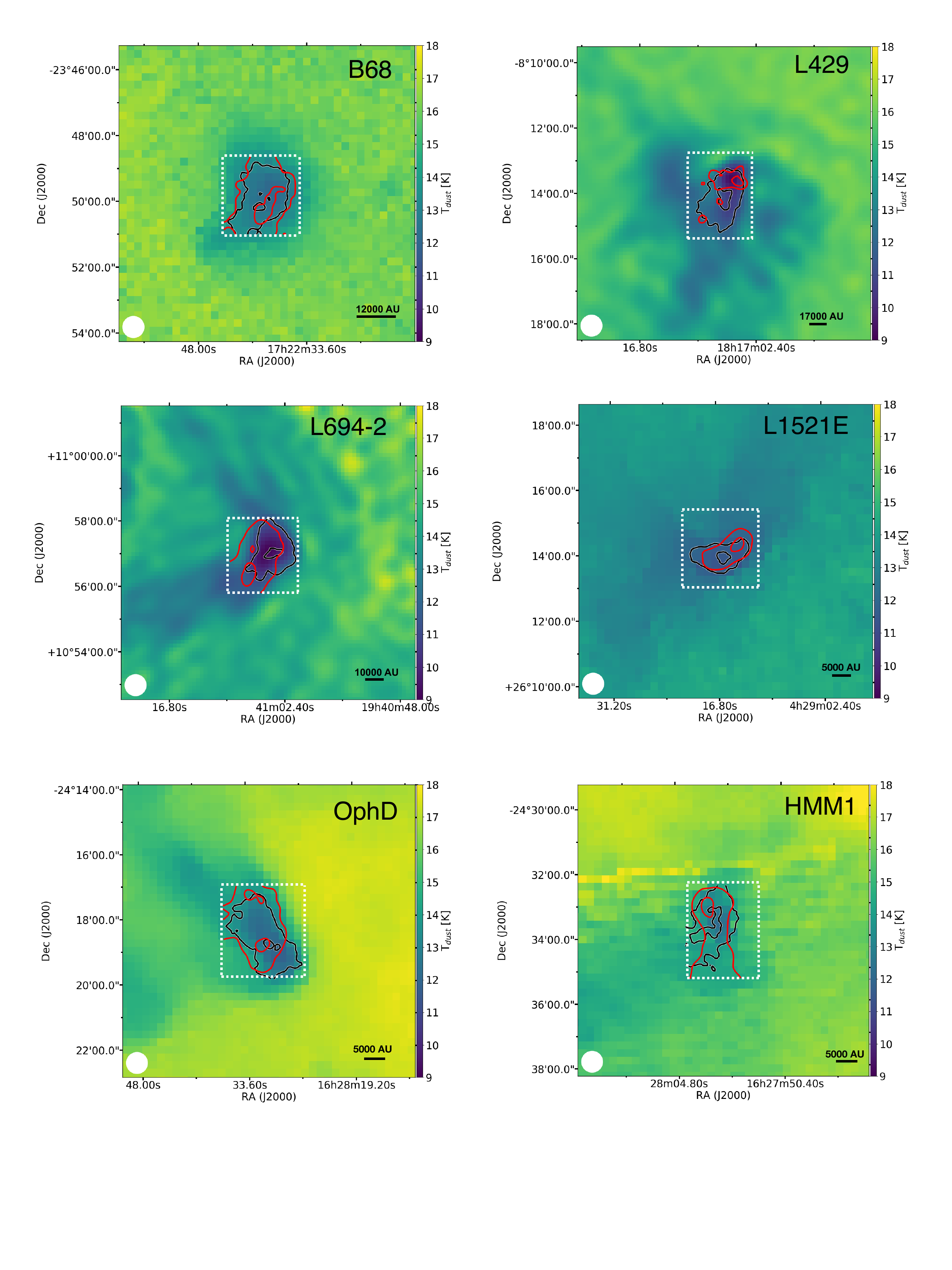}
 \caption{ T$_{dust}$ maps computed from {\em Herschel} SPIRE data. The red and black contours represent  90$\%$ and 50$\%$ of the integrated intensity peak of CH$_3$OH and $c$-C$_3$H$_2$, respectively. The dotted white box indicates the region mapped with the 30m telescope.
}
  \label{fig:Tdust}
\end{figure*}

The T$_{dust}$ maps of the cores in our sample were computed together with the H$_2$ column density from {\em Herschel}/SPIRE data at 250 $\mu$m, 350 $\mu$m, and 500 $\mu$m (see Figure~\ref{fig:Tdust}). For details see Section 2.3.

\section{Non-LTE effects on the N(CH$_3$OH)/N($c$-C$_3$H$_2$) abundance ratio}
\label{prova}
\begin{figure*}[h]
\centering
 \includegraphics [width=1.0\textwidth]{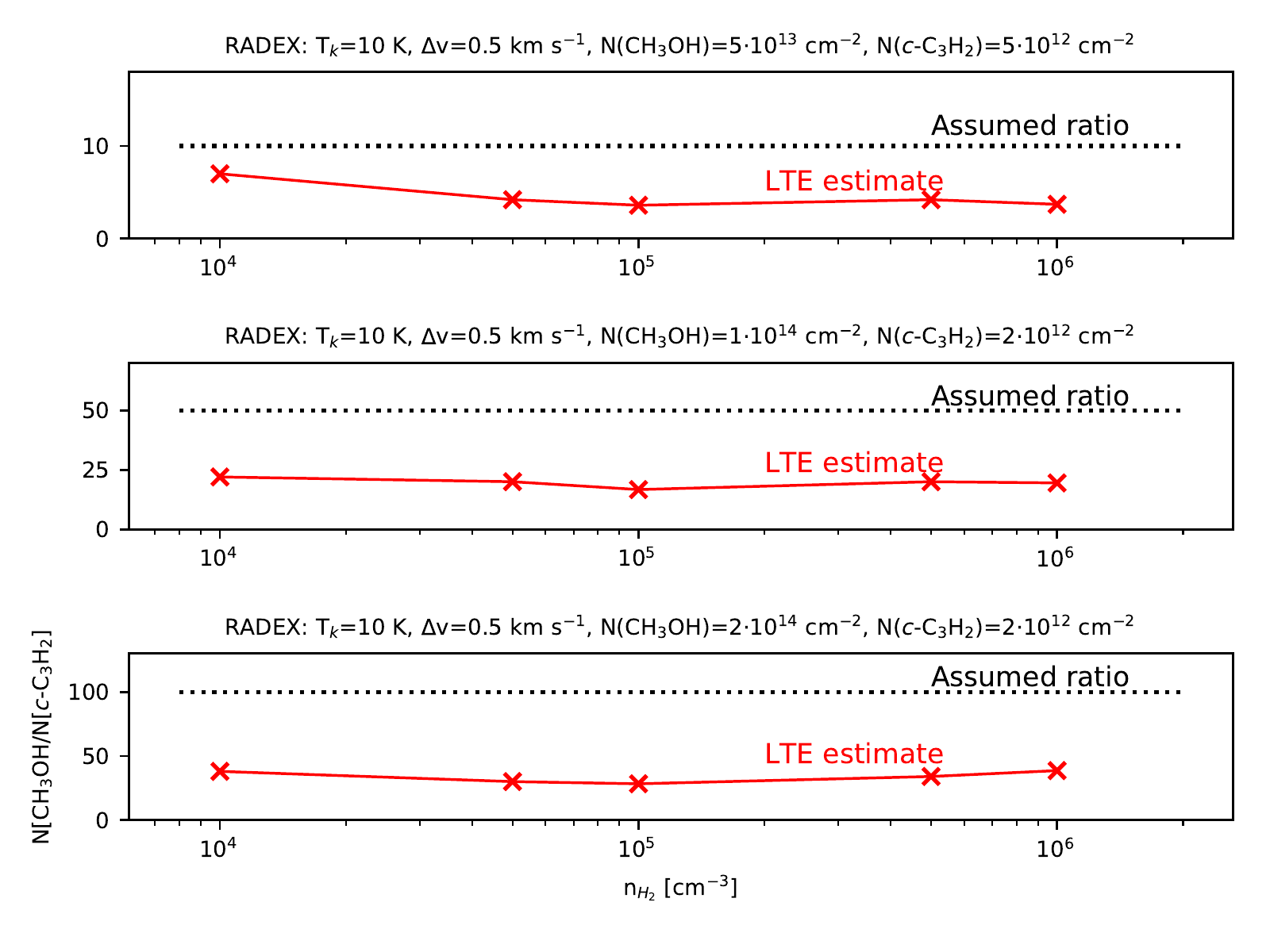}
 \caption{Variation of non-LTE effects on the N(CH$_3$OH)/N($c$-C$_3$H$_2$) abundance ratio at different volume densities. The intensities of the 2$_{1,2}$-1$_{1,1}$ ($E_2$) methanol line and the  $2_{0,2} - 1_{0,1}$ cyclopropenylidene line were calculated using RADEX at different volume densities, and assuming different ratios (shown as a horizontal dashed line in each panel). The parameters used for the RADEX calculations are shown at the top of each panel. The red crosses   show  the abundance ratios resulting from the column densities that have been calculated to reproduce the intensities derived with RADEX, assuming LTE and T$_{ex}$=10 K. }
  \label{fig:LTE_vs_RADEX}
\end{figure*}

Assuming the typical core conditions (T$_k$=10 K, $\Delta$v=0.5 km/s), three N(CH$_3$OH)/N($c$-C$_3$H$_2$) abundance ratios that cover the range shown in Figure~\ref{fig:profiles}, we ran RADEX for different densities between 1$\times$10$^4$ and 1$\times$10$^6$ cm$^{-3}$, covering the full range of densities in our sample. For each density and each abundance ratio, we looked at the RADEX prediction for the intensity of the 2$_{1,2}$-1$_{1,1}$ ($E_2$) methanol line and the  $2_{0,2} - 1_{0,1}$ cyclopropenylidene line. Using CASSIS and the LTE approximation with T$_{ex}$=10 K, we then estimated the CH$_3$OH and $c$-C$_3$H$_2$ column densities that reproduce the intensities derived from RADEX. Figure~\ref{fig:LTE_vs_RADEX} shows the intensity ratios calculated from the LTE approximation (red crosses) and the ratio assumed for the RADEX calculations (dashed horizontal lines). The LTE assumption underestimates the ratio by a factor of 2-3, but it does not show large variations within the volume densities covered. This means that the abundance ratio trends that we show in Figure~\ref{fig:profiles} are not going to be affected substantially by non-LTE effects.

\end{appendix}

\end{document}